\documentclass{iaus}
\usepackage{graphicx}

\title[~~The SEDs of Interacting Galaxies] 
{The SEDs of Interacting Galaxies}

\author[Lauranne Lanz et al.]   
{Lauranne Lanz$^1$, Nicola Brassington$^{1,2,3}$, Andreas Zezas$^{1,3}$, \\ 
Howard A. Smith$^1$, Matthew L. N. Ashby$^1$, Elisabete da Cunha$^4$, \\
Christopher Klein$^5$, Patrik Jonsson$^1$, 
Christopher C. Hayward$^1$,  \\
Lars Hernquist$^1$, \and Giovanni Fazio$^1$}

\affiliation{$^1$Harvard Smithsonian Center for Astrophysics \\ 
60 Garden Street, MS-10, Cambridge, MA 02138, USA \\ email: {\tt llanz@cfa.harvard.edu} \\[\affilskip]
$^2$School of Physics, Astronomy and Mathematics, University of Hertfordshire, \\
College Lane, Hatfield, AL10 9AB, UK\\
$^3$ Physics Department, University of Crete, P.O. Box 2208, 710 03 Heraklion, Crete, Greece\\
$^4$ Max Planck Institute for Astronomy (MPIA), K\"{o}nigstuhl 17, 69117, Heidelberg, Germany\\
$^5$ Department of Astronomy, University of California, Berkeley, CA 94720-3411, USA }

\pubyear{2011} 
\volume{284}  
\pagerange{1--12}
\jname{The Spectral Energy Distribution of Galaxies}
\editors{R.J. Tuffs \&  C.C.Popescu, eds.}
\begin{document}

\maketitle

\begin{abstract}
The evolution of galaxies is greatly influenced by their interactions. As part of a program to study interacting galaxies, we have measured and modeled the spectral energy distributions (SEDs) from the ultraviolet (UV) to the far-infrared (FIR).  We describe the constraints imposed on star formation histories by these SEDs, and the variations therein seen across the interaction sequence, and we compare the results of different star formation rate prescriptions applied to the data. The sample itself is based on the Spitzer Interacting Galaxy Survey (SIGS) of 111 galaxies in 50 systems, a project designed to probe a range of galaxy interaction parameters in the infrared. Our SEDs combine the Spitzer results with multiwavelength data from other missions,  in particular GALEX and Herschel.  The subset presented here is the sample for which FIR Herschel observations are currently publicly available.
\end{abstract}

\firstsection 
\section{The Spitzer Interacting Galaxy Survey}
\subsection{Sample Selection}

The SIGS sample is comprised of the Keel-Kennicutt (Keel et al. 1985) complete 
sample of interacting galaxies, which are chosen on the basis of association likelihood, supplemented with galaxies from their ``Arp sample''. Most of the systems within the complete sample are galaxies in early phases of interaction, so the inclusion of the Arp sample increases the range of interaction parameters covered. The Keel et al. (1985) selection criteria for the complete sample included a $\mid\Delta~\rm{v}\mid <$ 600~km~s$^{-1}$ requirement in order to exclude non-associated pairs. We imposed an additional distance constraint of cz $<$ 4000~km~s$^{-1}$ to retain high spatial resolution. These criteria yield a sample of 111 galaxies in 50 systems.

Previous samples tended to be chosen morphologically or based on infrared emission or optical emission line diagnostics. Due to the criteria used to select our sample, it spans a broad range of types of interactions and includes systems likely to be on their first pass as well as evolved systems. Similarly, in contrast to infrared-selected samples that tend to contain only the more active systems, our sample covers a large range of degrees of nuclear 
and/or starburst activity. Finally, our use of infrared diagnostics, rather than optical, significantly reduces the effects of obscuration.

\subsection{Observations}

The sample has a complete set of mid-infrared (MIR) observations taken by the Spitzer Space Telescope, including photometry with both the Infrared Array Camera (IRAC; Fazio et al. 2004) and the Multiband Imaging Photometer (MIPS; Rieke et al. 2004) as well as nuclear spectroscopy with the Infrared Spectrograph (IRS; Houck et al. 2004). In addition, we have $>90$\% coverage in the near-ultraviolet (NUV) and far-ultraviolet (FUV) with Galaxy Evolution Explorer  (GALEX; Martin et al. 2005).  Presently, $\sim$20\% of the galaxies also have publicly available FIR photometry from the Spectral and Photometric Imaging Receiver (SPIRE; Griffin et al. 2010) aboard the Herschel Space Observatory. Together, these infrared and ultraviolet observations  give a view of both the obscured and unobscured star formation. We supplement these data with near-infrared observations from the Two Micron All Sky Survey (2MASS; Skrutskie et al. 2006), far-infrared photometry from the Infrared Astronomical Satellite (IRAS; Neugebauer et al. 1984), and UBV photometry from the Third Reference Catalogue (de Vaucouleurs et al. 1991). We are in the process of obtaining additional UV data from Swift Satellite imaging (Roming et al. 2005). Figure 1 shows  a subset of these observations for a typical sample pair of galaxies, NGC 3430 (top) and NGC 3424 (bottom).

The apparent extent of a galaxy can vary significantly between the ultraviolet and the infrared. Therefore, we used S-Extractor to determine the optimum aperture size in both NUV and 3.6 $\mu$m images. The larger aperture was then consistently applied to all the observations when measuring photometry. Uncertainties in the measured fluxes consist of a sum in quadrature of the calibration and Poisson noise values. 

\begin{figure}[t]
\begin{center}
 \includegraphics[width=\linewidth]{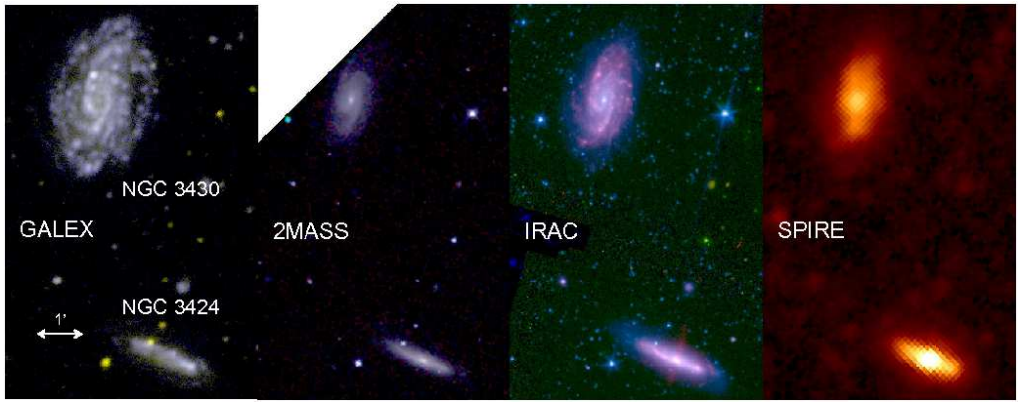} 
 \caption{The interacting spiral pair of NGC 3430 (top) and NGC 3424 (below) as observed by GALEX (NUV in yellow and FUV in blue),  2MASS (J in blue, H in green, and  K$_{s}$ in red), IRAC (3.6 $\mu$m in blue, 4.5 $\mu$m in green, and 8.0 $\mu$m in red), and SPIRE 250 $\mu$m. These galaxies are in an early interaction stage and show some of SIGS's range in relative UV-to-IR morphology, exemplified by the contrast of NGC 3430's extended UV disk and NGC 3424's 
consistent morphology.}
   \label{fig1}
\end{center}
\end{figure}

\section{Spectral Energy Distributions}

Fig. 2 presents two typical SEDs from the subsample: NGC 3430, part of a spiral galaxy pair and M51B, a dwarf galaxy interacting with M51A. These SEDs demonstrate some of the variety observed. To extract quantitative parameters such as the star formation rate and stellar and dust masses, we used MAGPHYS (da Cunha et al. 2008), which uses stellar libraries from Bruzual \& Charlot (2003) in combination with infrared dust spectral libraries to fit the SED such that energy absorbed in the UV and optical is re-absorbed in the infrared. This code provides posteriors of such parameters as temperatures of warm (30-60 K) and cold (15-25 K) dust, stellar and dust masses, and star formation rate (SFR). We find good agreement between the dust luminosity determined by this method and the integrated 8-1000 $\mu$m luminosity.
 
\begin{figure}[h]
\begin{center}
 \includegraphics[width=0.8\linewidth]{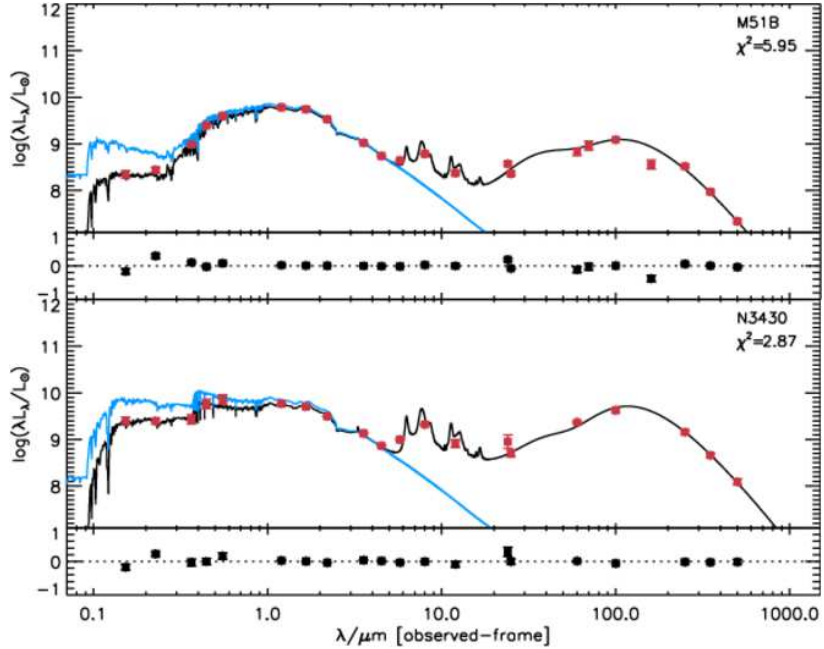} 
 \caption{SEDs of M51B and NGC 3430, showing some of the range in SED shape seen, particularly in terms of the relative 
strengths of the MIR to the optical and the UV to the optical, overplotted with the fits from MAGPHYS (da Cunha et al. 
2008) showing the best fit model (black) and the unattenuated stellar fitted spectrum (blue).}
   \label{fig2}
\end{center}
\end{figure}

\section{Results}
\subsection{Variations with Interaction Stage}

Since it is notoriously difficult to precisely identify the interaction stage of any one galaxy system, we use the Dopita et al. (2005) five-stage classification scheme, where stages 2-4 represent mildly, moderately, and strongly interacting systems. Fig. 3 shows the warm (30-60K; blue) and cold (15-25K; green) dust temperatures (left) and specific star formation rate (sSFR; right) derived by MAGPHYS for the 22 galaxies in the 10 systems with currently available SPIRE observations with the mean and standard deviation of each set over-plotted. The mean specific star formation rate does not appear to show any significant variation as the interactions evolve in this dataset. While the cold diffuse dust likewise does not increase in temperature with interaction stage, the warm dust, associated with both interstellar medium and the birth clouds, shows a probable increase with interaction evolution.

\begin{figure}[t]
\begin{center}
 \includegraphics[width=\linewidth]{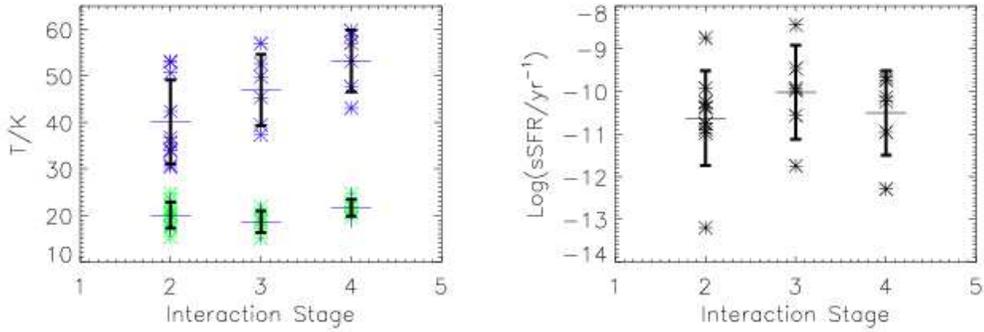} 
 \caption{Specific star formation rates (left) and warm (blue) and cold (green) dust temperatures (right) plotted against the Dopita et al. (2005) interaction stage, showing that based on the first 22 galaxies, the only one of these three parameters to vary significantly with interaction stage is the temperature of the 30-60 K dust associated with birth clouds and the ISM.}
   \label{fig3}
\end{center}
\end{figure}

\subsection{Comparison of SED-derived sSFR with sSFR prescriptions}
Future galaxy surveys, particularly those studying objects at high redshirt, will probably not have the opportunity or the capacity to exploit the same wealth of observations we used for this sample. As such, it is important to understand which one- or two-band SFR prescriptions work best for interacting systems. In Fig. 4, we compare several indicators based either solely on UV or IR emission or a sum thereof to the sSFR derived from MAGPHYS. We generally find that the SED-derived sSFR agrees better with a prescription that includes contribution from the IR. However, NGC 3430 is a counter-example: its extended UV disk suggests that the IR-dependent prescriptions, which generally assume optical thickness, miss a significant portion of the recent star formation. We are currently extending this work to include all 111 galaxies in the SIGS sample, allowing us to thoroughly test the sSFR indicators.

\begin{figure}[h]
\begin{center}
 \includegraphics[width=0.6\linewidth]{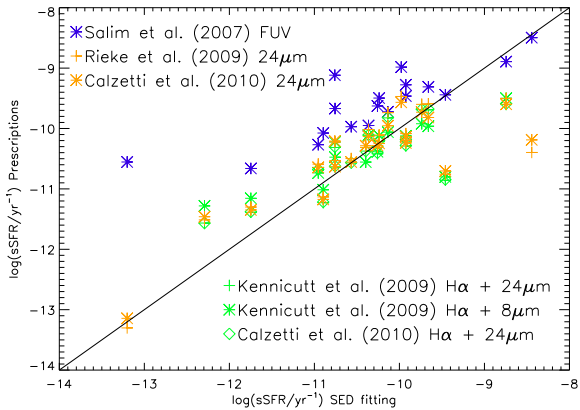} 
 \caption{Comparison of specific star formation rates derived by MAGPHYS from fitting the SED to those calculated using one- or two-band prescriptions. The two objects with the highest SED-derived sSFR both have extended UV disks.}
   \label{fig4}
\end{center}
\end{figure}

\noindent
\textbf{Acknowledgements}

We are grateful to Don Neill and Susan Neff for their assistance with the GALEX data and to 
Volker Tolls and Joe Hora for their assistance in reducing Herschel SPIRE observations.
This work was supported by the Harvard College Observatory and NASA through an award issued by JPL/Caltech via contract \#1369566.

\end{document}